# Mechanistic Data Science for Modeling and Design of Aerospace Composite Materials

**Satyajit Mojumder** [1]
Graduate Research Assistant
Theoretical and Applied Mechanics Program
Northwestern University
Evanston IL-60208, USA
Email: satyajit@u.northwestern.edu
ASME Graduate Student Member

**Lei Tao[1]**
Postdoctoral Research Associate
Department of Mechanical Engineering
University of Connecticut
Storrs, Connecticut 06269, USA
Email: lei.tao@uconn.edu

**Ying Li** [2]
Assistant Professor
Department of Mechanical Engineering
Associate Member, Polymer Program, Institute of Materials Science
University of Connecticut
Storrs, Connecticut 06269, USA
Email: ying.3.li@uconn.edu

**Wing Kam Liu[2]**
Walter P. Murphy Professor and Co-founder of HIDENN-AI, LLC
Department of Mechanical Engineering
Northwestern University
Evanston, IL-60208, USA
Email: w-liu@northwestern.edu

## ABSTRACT

*Polymer matrix composites exhibit remarkable lightweight and high strength properties that make them*

*attractive for aerospace applications. Constituents' materials such as advanced polymers and fibers or fillers*

---

[1] Equal contribution.
[2] Corresponding author.

*with their hierarchical structure embed these exceptional properties to the composite materials. This hierarchical structure in multiple length scales provides an opportunity for designing the composite materials for optimized properties. However, the high dimensional design space for the constituents' materials and architectures choice of the composites makes it a challenging design problem. To tackle this high dimensional design space, a systematic, efficient approach named mechanistic data science framework is proposed in this work to identify the governing mechanisms of materials systems from the limited available data and create a composite knowledge database. Our composite knowledge database comprises the knowledge of polymers at the nanoscale with nano reinforcement, the unidirectional structure at the microscale, and woven structure at mesoscale mechanisms that can be further used for the part scale composite design and analysis. The mechanistic data science framework presented in this work can be further extended to other materials systems that will provide the materials designer with the necessary tools to evaluate a new materials system design rapidly.*

**INTRODUCTION**

Composite materials, having unprecedented and unique properties to their constituents' materials, are replacing many conventional materials applications in the automotive, aviation, and space industry [1,2]. With the lightweight, higher stiffness and strength, less corrosion and fatigue, it is a promising material for aerospace applications. The aerospace composite materials are now in practice for commercial passenger aircraft to military aircraft, even the spacecraft application [3]. Different composites are used in the aerospace industry, where polymer matrix composites have a major share due to their lightweight and high strength properties. For example, the most recent Boeing 787 aircraft comprises 80% composite materials by volume, 50% composite by weight with 20% aluminum, 15% titanium, 10% steel, and 5% others [4]. Though polymer matrix composites have remarkable properties, it is still not suitable for high-temperature applications [5–7]. Therefore, significant research efforts have been made to address it in recent years.

Polymer matrix composites have two major constituents: the polymeric materials as the matrix phase, and some reinforcing materials in the form of fiber or fillers. Combining the properties of the fiber and filler with the base polymers, exceptional improvement in the mechanical properties can be achieved. However, for aerospace applications, very few polymeric materials have been explored to date [8,9]. The most promising candidate in the polymer is the epoxy resin that has been used extensively with carbon, glass, Kevlar, boron fibers, and so on. Though epoxy-based polymers impart excellent mechanical properties into the composites, they are a thermosetting polymer that is not recyclable. In terms of thermoplastics, they are recyclable, and polypropylene, polyetheretherketone (PEEK), Nylon 66, polymethylmethacrylate (PMMA) are the name of few thermoplastics that have been extensively investigated with a lower applicable temperature range. With this need in mind, novel thermoplastic materials are under investigation that can address this significant challenge of high-temperature application to be a successful candidate for the aerospace application [9]. In terms of new polymeric materials development, the chemical design space of the polymers needs to be considered for the synthesis of high-temperature polymeric materials. The exploration of the almost infinite chemical space of polymeric materials combines two sub-tasks: the generation of new polymers to populate an entire chemical space, and the establishment of accurate and robust machine learning models to evaluate the properties in the chemical space.

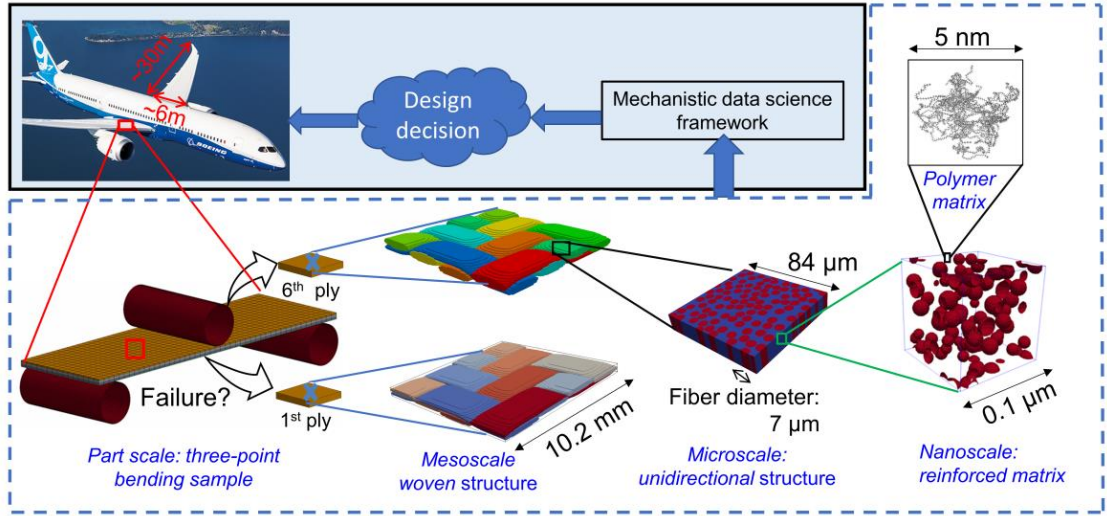

**Fig. 1**. Aerospace composite materials and its structure in nanoscale as reinforced matrix, microscale as unidirectional structure, mesoscale as woven structure and a part scale three-point bending sample.

In composite form, the polymer and the reinforcing phase of the materials form a hierarchical structure that shows multiple length scales [10,11]. The architectural design of the composite materials is the key to their structure and load-bearing properties [12]. As shown in **Fig. 1**, a composite part can have woven [13,14] or braided structure [15] in the mesoscale, and further zoom-in of yarn of the woven structure reveals the unidirectional (UD) structure [16,17] of the composite materials. The matrix phase of the materials is a polymer that can be reinforced with nanoparticles to improve the toughness of the polymers [18–22]. Novel reinforcement at the nanoscale can change matrix material's mechanical properties and alter the thermal properties by changing the glass transition temperature of the composite. Recently nanotubes [23], nanoparticles [24], and nanoplatelets [25–28], such as $MoS_2$ has been studied as the nanoscale reinforcement of the matrix materials that provide excellent mechanical and fracture properties. For the design of the real composite part, several design variables for multi-scale architectural composite structures can be identified. For instance, at the mesoscale woven architecture, the yarn fiber volume fraction, yarn angle, yarn geometries, weave patterns, etc., can be designed for better materials performance [14,29]. Similarly, at the microscale unidirectional design, fiber volume fraction can be altered [30]. At the nanoscale, the reinforcement volume fractions can be varied to change the toughness and other properties [31] of the matrix materials. The chemical elements and connectivity of the polymeric matrix can be substituted to possess various structural features and properties. Considering all these design parameters, the composite material system becomes very high dimensional, which requires extensive experiments and computational efforts to characterize the different designs of the materials architectures.

Due to the high dimensional design space of composite architecture at multiple length scales, it becomes a challenging problem from the materials design point of view, where the properties can be optimized in terms of all these design variables. Recent advancements in data science techniques and its capabilities to solve high dimensional design problems make it a suitable candidate for exploring [32–34] this high dimensional composite design problem. However, most of these data science techniques are data-hungry (requires a lot of experimental or modeling data), and very limited data are available for the composite materials [35]. Mechanicians address the composite materials problem in a deterministic way, where they can analyze the properties and performance for a given architecture and understand their mechanics. With this incomplete scientific knowledge and the limited available data, the design problems become intractable. Still, there is a dire need to assess different materials design rapidly for a part design from the materials designer community. This data-driven research effort has been gaining significant attention in recent years. Different machine learning techniques are applied to characterize and predict composite material performance at different length scales. Using the data-driven approach, surrogate models are proposed for learning the constitutive relations and predicting the elastic properties of the materials [36,37]. These surrogate models used different machine learning techniques such as artificial neural networks (ANN) [36,38,39], Gaussian process regression [40], logistic regression [41], support vector machines, and so on. Also, a convolutional neural network (CNN) is used to evaluate the composite design and performance prediction [42–44]. Besides, a recurrent neural network (RNN) [45,46] is adopted to predict plasticity for composite microstructures. Physics informed neural network (PINN) [47] has also been proposed to predict the response of synthetic composite material. The deep learning algorithm further extends to damage and structural health monitoring applications [39,48] for composite materials. All these methods require a massive amount of data for the training purpose and do not work well outside the training range. Another effort is employed by researchers to develop reduced-order models for faster estimation of the composite response. Liu et al. proposed the self-consistent clustering analysis [49,50], which is a mechanistic reduced-order model that can predict the homogenized response for the composites. It is further explored for multiresolution clustering analysis [51] and applied for multi-scale composite materials modeling. Few other data driven techniques such as FEM clustering analysis (FCA) [52], virtual clustering analysis (VCA) [53], MAP123 [54–56], hierarchical deep learning neural network (HiDeNN) [57] has been proposed to predict the homogenized response of the composite materials. While these methods are successful to predict the composite response at different scales, their combination can provide complete analysis and design tool for the composite materials.

To address the challenge of the data scarcity in the composite materials study, a systematic approach is needed to combine the existing knowledge with the extracted mechanism by intelligently mining the limited available data. This approach is termed Mechanistic Data Science (MDS), where existing data science tools are employed to extract the governing mechanism from a system. Liu et al. recently proposed this general MDS framework [58] and showed its applications in different fields of science and engineering. In this work, we took this systematic approach and applied it to the composite materials system design. MDS approach has several modules to manage the data for a system efficiently. It starts with a multi-modal and multi-fidelity data generation and collection, followed by mechanistic feature extraction, and further reduced by dimension reduction techniques. Once the governing features are identified, reduced-order surrogate models and mechanistic learning through regression and classification can establish an easier way to determine the relation of the governing parameters and properties of the system. Once the governing mechanisms are understood, they can be extended to new materials systems and designs. Throughout this process, a knowledge database is created that can be used to design and analyze a new system. In this work, we approach the multi-scale composite modeling as shown in **Fig. 1** using the MDS framework and discussed key steps for composite knowledge database creation.

This work is organized as follows. **Section 2** provides the MDS framework for integrated multi-scale composite materials design and analysis. **Section 3** discusses the polymer design, reinforced matrix design at the nanoscale, unidirectional composite design at the microscale, woven composite design at the mesoscale, and composite knowledge database creation for part scale design. Finally, the perspective on MDS-based multi-scale aerospace composite materials and conclusions are discussed in **Section 4**.

**MECHANISTIC DATA SCIENCE FRAMEWORK**

Mechanistic data science (MDS) framework is a systematic approach that can be applied to analyze a materials system combining the existing scientific knowledge and identifying governing mechanisms from the available data. In principle, the MDS approach uses data science tools to extract information from the existing data to build a knowledge database of the system for understanding the scientific principle of the system. Typically, in engineering, we encounter three types of problems: i) limited physics problem with abundant data, ii) incomplete physics problem with limited data, iii) known physics but computationally expensive. Using the MDS framework, these three types of problems can be approached. For instance, with the first type of problem, MDS can find the governing mechanisms from the data and help understand the physics of the problem. For the second type of problem, it can fill the gap for incomplete physics by analyzing the limited amount of data available. For the computationally expensive problem, it can help solve the computationally intractable problems through the data-driven computational approach. MDS framework has six different modules that interact to build the system knowledge database that can be further used to design, analyze, and optimize the system. MDS is a general framework that can be applied to a wide range of the physical problem. Here we discussed it in the context of the aerospace composite materials and their multi-scale modeling. Therefore, the six modules are introduced below with the necessary steps to model and design the aerospace composite materials.

**Multimodal Data Generation and Collection**

MDS relies on the existing data to identify the governing mechanism of a physical system. Therefore, data collection and generation are the starting point to analyze any system. Though abundant data are available in finance and other fields, very limited data are available for engineering problems. In many cases, the experiments are very expensive, and the existing dataset is incomplete. Synthetic data through numerical modeling and simulations can help this data-scarce problem. Also, the data can come from multi-modal sources such as experiments, imaging, sensing, modeling, and simulations. Processing data from this multi-modal source is a significant challenge. Also, data can be of multi-fidelity (different resolutions). Several data science tools are available to preprocess the data and prepare it for further analysis, such as data wrangling, cleaning, augmentation, etc. Data visualization is also a powerful tool to identify the outliers and patterns of the data.

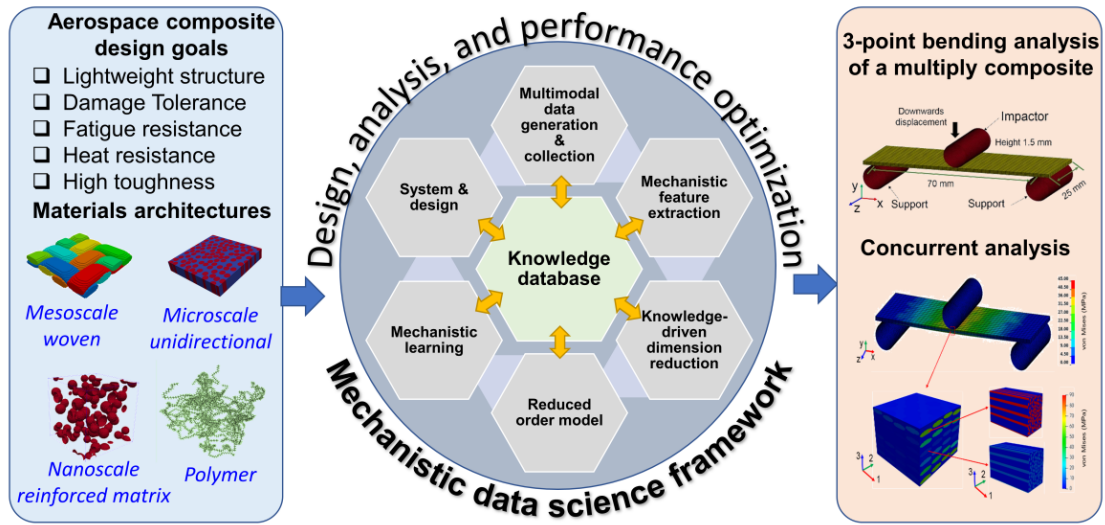

**Fig 2.** Mechanistic data science framework and its six components to create knowledge database for composite design, analysis and performance optimization.

For aerospace composite materials, very limited experimental data are available. To date, very few polymers are tested as the matrix materials of the composites. Also, the composite has a vast design space in nano, micro, meso, and part scales, as illustrated in **Fig. 1**. which is impossible to test through physical or numerical experiments. With the limited data and scientific knowledge available in the composite materials, we can perform simulations to generate data to identify the mechanisms in the design space. One challenge to generating data for composite materials is its multi-scale nature.

**Mechanistic feature extractions**

After the data generation and collection process, mechanistic features are extracted from the data. These techniques provide an efficient way for data management by discarding unnecessary features. Tools such as Fast-Fourier Transform (FFT), Short-time Fourier Transform (STFT), Wavelet Transform, etc., can be used to extract mechanistic features from the data. Also, meaningful mechanistic features obtained from a physical simulation such as strain concentration tensor, stress-distribution can be used as a mechanistic feature. For example, for the aerospace composite materials system shown in **Fig. 1**, the mechanistic features such as elastic, yield, and hardening properties are extracted from the stress-strain curves for the reinforced matrix and unidirectional composites.

**Knowledge-driven Dimension Reduction**

In the dimension reduction step, high dimensional data or mechanistic features of the system are reduced to low dimensional space. Principal component analysis (PCA) and clustering analysis (unsupervised learning) are two widely used techniques for the dimension reduction of the data. Also, scaling and dimensional analysis of the data can provide nondimensional numbers with physical intuition into the problem [59,60]. Several features can be combined into a new feature to reduce the dimension of the features. For instance, the fiber density, matrix density, and fiber volume fraction can be reduced to a single feature, composite density, using a well-known rule of mixtures formula.

**Reduced-Order Model**

Reduced-order model aims to reduce computation needed to solve a physical problem and provide an inexpensive solution to the problem. Often the reduced-order models are not robust to parametric changes and need the precomputation of the parametric space to be predictive in real-time. Mechanistic based reduced-order model is based on the governing mechanism of the problem that makes it predictive for the parameter range not precomputed. Self-consistent clustering analysis (SCA) [49] used in this work for the stress-strain response prediction is such a mechanistic reduced-order model. In SCA, the elastic response of the materials is precomputed in an offline step following a dimension reduction clustering step and interaction tensor calculation of the clusters. In the online stage, the mechanistic Lipmann-Schwinger equation is solved for the cluster to predict the homogenized stress-strain response. As the cluster reduces the computational degree of freedoms drastically (from millions of voxels to a hundred clusters), a significant computational speedup is attained that enables us to perform many tensile simulations for a wide choice of the microstructures in a short time and build the knowledge database.

**Mechanistic Learning**

Mechanistic learning is used for the regression and classification of the mechanistic features. Once the features are identified, the input and output for the system can be related using the regression tool. A wide choice of regression tools is available in the data science techniques. A neural network is one of the most used tools due to its universal approximation capability of any function. Previously, constitutive relations were predicted by several researchers for the composite materials using neural networks and other regression techniques. For the multi-scale analysis using neural network, materials properties of the reinforced matrix and fibers can be used to obtain the unidirectional composite stiffness properties that are further used to model the yarn properties of the woven composite. A neural network can learn the mechanistic relation of the unidirectional composite properties as a function of the reinforced matrix, fiber, and fiber volume fractions.

**System and Design**

System and design identify the design parameters for the system and try to investigate the system response as the design parameters get changed. Eventually, all these changes are stored in the composite knowledge database. The extracted relation for one materials system can be easily transferred to a different materials system or a new system design for the same material's constituents. For aerospace composite materials, the design space is relatively high and distributed across multiple scales. For instance, a wide choice of polymers is available to design the composites. In nanoscale design, the volume fraction and the distribution of the nanofiller can be varied. Similarly, for microscale, the fiber volume fractions, orientations can be designed, and for mesoscale, the shape and size of the yarn, yarn fiber volume fractions, yarn angle, and woven pattern can be designed for a specific part scale application. Considering all these design choices at different length scales, the materials system has many parameters to optimize for a specific application. Once the knowledge database is created for these different design parameters, they can be readily utilized for any composite part design analysis and optimization.

**COMPOSITE KNOWLEDGE DATABASE CREATION**

Composite knowledge database consists design of the material's level as well as the material's architecture at different length scales. We first discuss the design of matrix materials with a focus on high-temperature polymer design. Further, we show the MDS steps at nano-, micro-, meso- length scales for the reinforced matrix, unidirectional composite, and woven composite modeling, respectively.

**Polymer Design**

Various polymeric materials - as the continuous phase of the composite - have been wildly used with different discontinuous reinforcements. The thermo-mechanical property of the polymer matrix plays a significant role in composite properties. For example, high-temperature conditions mostly affect the polymer matrix other than the reinforcing fillers, and the compression behavior of the composite is primarily controlled by the polymer matrix as well. Therefore, besides the common low-cost polymer matrices like polyesters and vinyl esters, more attention has been focused on polymer matrices with desired properties like high glass transition temperature, thermal stability, or toughness. With the increasing amount of polymer databases, data-driven regression methods such as Gaussian process regression arise to build correlations between polymer chemical structures and their thermo-mechanical properties. Moreover, data-driven design methods such as generative machine learning models provide more opportunities for the molecular design of high-performance polymers.

In the largest database of experimentally reported polymers, "PolyInfo" [61] from Japan's National Institute for Materials Science, the chemical structures of more than 13,000 homopolymers are available, and 36 different properties are collected. Other useful databases also include "PubChem", "Polymer Property Predictor and Database", and "NIST Synthetic Polymer MALDI Recipes Database" from the United States. To utilize polymer databases in any data-driven design and evaluation methods, the first and most crucial step is the proper digitalization of polymers. The International Union of Pure and Applied Chemistry (IUPAC) nomenclature of organic chemistry refers to polymers in a human-readable way, like poly(non-1-ene). It can be equally denoted with the Simplified Molecular Input Line Entry System (SMILES) notation, like '*C(C*)CCCCCCC'. Then cheminformatics packages like RDkit and Pymatgen can process the polymer structures into numerical fingerprints or descriptors. A machine-learning model becomes applicable only when the numerical representation of polymers is appropriately obtained. It is worth noting that three main steps are involved in establishing a machine-learning model: 1. determining a proper structure representation of polymers (based on monomer, repeat unit, or oligomer, etc.). 2. Utilizing an appropriate feature engineering (fingerprint, descriptor, or molecular graph, etc.). 3. Implementing a proper machine-learning algorithm (linear regression, feed-forward neural networks, or Gaussian process regression, etc.).

With a careful digitalization of polymer structures, establishing a machine-learning model to correlate the polymer structure-property relationship properly is essential to all data-driven design strategies. The most commonly used method for polymer discovery is based on high-throughput screening. When a machine-learning model is well trained on existing polymers, it can recognize critical structural features that control the material properties. For example, as shown in **Fig. 3**, focusing on the discovery of high glass transition temperature ($T_g$) polymers, there are ~7,000 existing polymers with measured $T_g$ in the PolyInfo dataset, and ~2000 of them have $T_g$ > 200 °C. The training of a machine-learning model on this dataset leads to a function that takes in the structure of a polymer and then outputs an estimation of its $T_g$. Applying the obtained machine-learning model on newly proposed polymers can evaluate their properties much more efficiently than the time-consuming experimental measurements. High-throughput screening of 1 million proposed hypothetical polymers discovers more than 65,000 promising candidates with $T_g$ > 200°C [9]. The design strategy of using high-throughput screening requires a well-established machine-learning model. Similarly, the design methods of using genetic algorithms and combinatorial enumerations generate new polymers in different ways. Still, ultimately all the candidates need to be evaluated by a reliable and accurate machine-learning model.

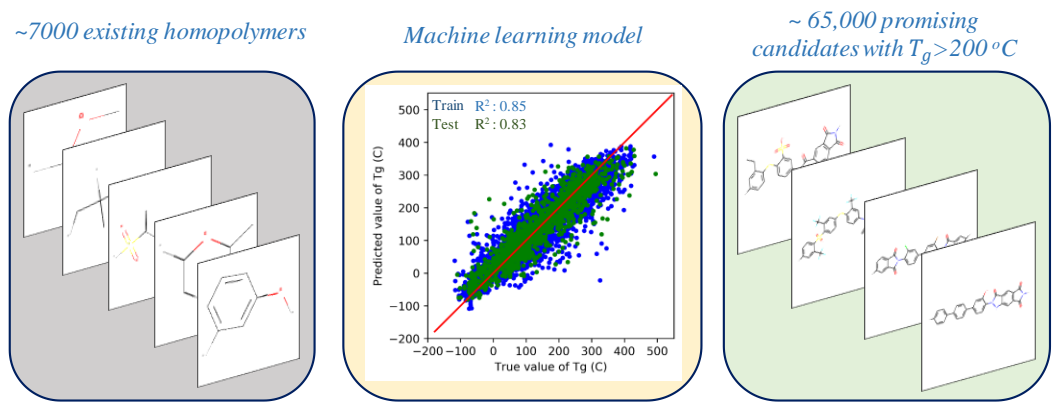

**Fig 3**. Discovery of new high-temperature polymers. A DNN machine learning model is first trained on a dataset of ~7000 existing polymers. Then, the model is used to screen 1 million hypothetical polymers to discover ~65,000 promising high-temperature structures whose $T_g$ > 200°C.

The generation of polymer candidates is usually followed by the evaluation of their properties, and most of the generated candidates are invalid or undesired structures, with only a small portion found promising. Therefore, the design process using a predictive machine-learning model can be tiresome sometimes. To improve the design productivity, the generative machine-learning model integrates extra constraints into the method. The variational autoencoder (VAE) imposing chemical grammar can alleviate the generation of syntactically invalid structure [62]. And the generative adversarial networks (GANs) [63], along with reinforcement learning, can bias the polymer generation towards the direction with desired properties (**Fig. 4**). Such a design strategy of using a generative model limits the design space to only reasonable structures. In the meantime, it is not restrained by pre-defined rules for molecular generations. The generation of millions of valid data points is achievable for polymer data-driven analysis. However, it is still a relatively small data volume considering that small molecules has beyond the size of billions. To better cover the ideally infinite chemical space of polymers, the more valid candidates the better. In addition, since a newly designed polymer matrix are expected to have multiple high thermo-mechanical properties, multi-task learning for predictive machine-learning model or multi-reward/penalty for generative machine-learning model will attract more attentions. The data-driven design of high performance polymer matrices will lay a solid foundation for the data-driven design of aerospace composite materials.

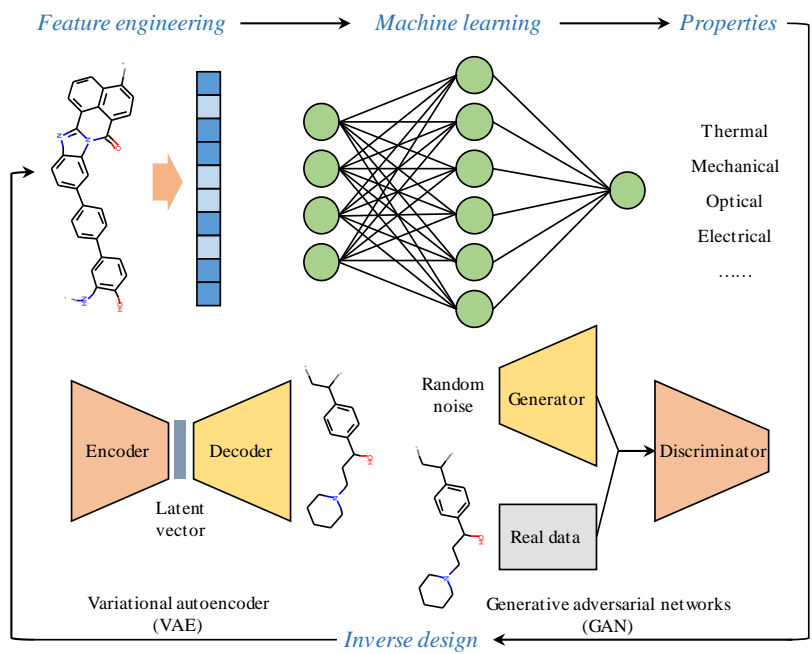

**Fig 4.** The forward prediction of polymer properties and the inverse design of polymer structures. Variational autoencoder (VAE) and generative adversarial networks (GAN) generate new polymer structures, whose properties can be evaluated by the forward prediction machine-learning model.

**Nanoscale Reinforced Matrix design**

Polymer is the base material for the polymer matrix composites; however, nanoparticles are often added as a reinforcement in the matrix phase. The toughening and thermal properties of the composite can be enhanced by adding these nanoparticles to the polymer matrix. Typically, the mechanical properties increase with nano-size (<500 nm) inclusions up to a certain volume fraction; after that, the particles tend to agglomerate together and show poor mechanical properties. Here we chose epoxy as the base matrix and added up to 15% silica inclusions, a typical nanoparticle to increase the toughness of epoxy [19]. The silica particles have an average diameter of 20 nm with a maximum of 30 and a minimum of 10 nm. The structure of the silica-reinforced epoxy matrix is shown in **Fig. 5(a),** and their stress-strain response for different volume fractions of the silica particles is shown in **Fig 5 (b)**. As the volume fraction of silica is increased, the elastic properties of the reinforced epoxy are enhanced. Also, the temperature-dependent stress-strain behavior is shown in **Fig 5(c)** shows that for a specific temperature range, the silica inclusions can impart maximum toughness. The volume fractions of the silica particles and the temperatures are considered two important design parameters for the nanoscale reinforced matrix.

As the first step of the MDS process, we start with the experimental microstructures of the reinforced matrix materials. The microstructures are characterized based on the particle size and the volume fraction, and reconstructed as a representative volume element (RVE) for the nanoscale. The stress-strain simulations are performed using a mechanistic reduce order model, Self-consistent Clustering Analysis (SCA). For the brevity of this article, the details of the SCA are not discussed here, and the interested readers are referred to the literature [49,50]. Using SCA, we first compute the elastic response of the reinforced epoxy structure. Based on that, an unsupervised learning algorithm (K-means clustering) is applied to group the voxels of the nanostructure. Later, the interaction tensors for the clusters are computed and stored as an offline database. In the online stage, the clusters are solved using mechanistic Lipmann-Schwinger equations to predict the nanostructure's stress—strain response. The benefits of using SCA here are twofold: i) it provides an efficient and accurate way to extract homogenized properties of the nanostructure for any loading conditions, (ii) the computation speedup is significant (~5000 times for UD RVE [16]) that allow us to explore the design space of the composite materials.

As a next step of the MDS, the reduced-order model (SCA) generated stress-strain reinforced epoxy is further analyzed. Several mechanistic features such as elastic modulus ($E$), yielding strain ($\varepsilon_{elas}$), yield stress ($\sigma_y$), hardening parameters ($K, n$) can be extracted to represent the essential features for a particular stress-strain curve. Once these mechanistic features are identified, they can be used to regenerate a stress-strain behavior of the reinforced matrix as needed. This reinforced matrix is used as a matrix material in the larger length scales, such as unidirectional and woven composites. Considering the hardening and damage of the reinforced matrix phase, which differs significantly from the reinforcing materials, their amounts are important for an accurate composite response. The process of extracting these mechanistic features is shown in **Fig. 5(d)** for a sample stress-strain curve. The elastic modulus is extracted from the slope of the linear portion of the stress-strain curve, and a 0.2% strain offset method is used to identify the yielding strain and yielding stress. Hollomon strain hardening equation ($\sigma = K\varepsilon^n$) is used to describe the elastoplastic behavior of the materials, where $K$ is the strength coefficient, and $n$ is the strength hardening exponent [64]. These mechanistic features are stored in the knowledge database for silica volume fractions in the range of 0-15% by volume and temperature range of 213-295K.

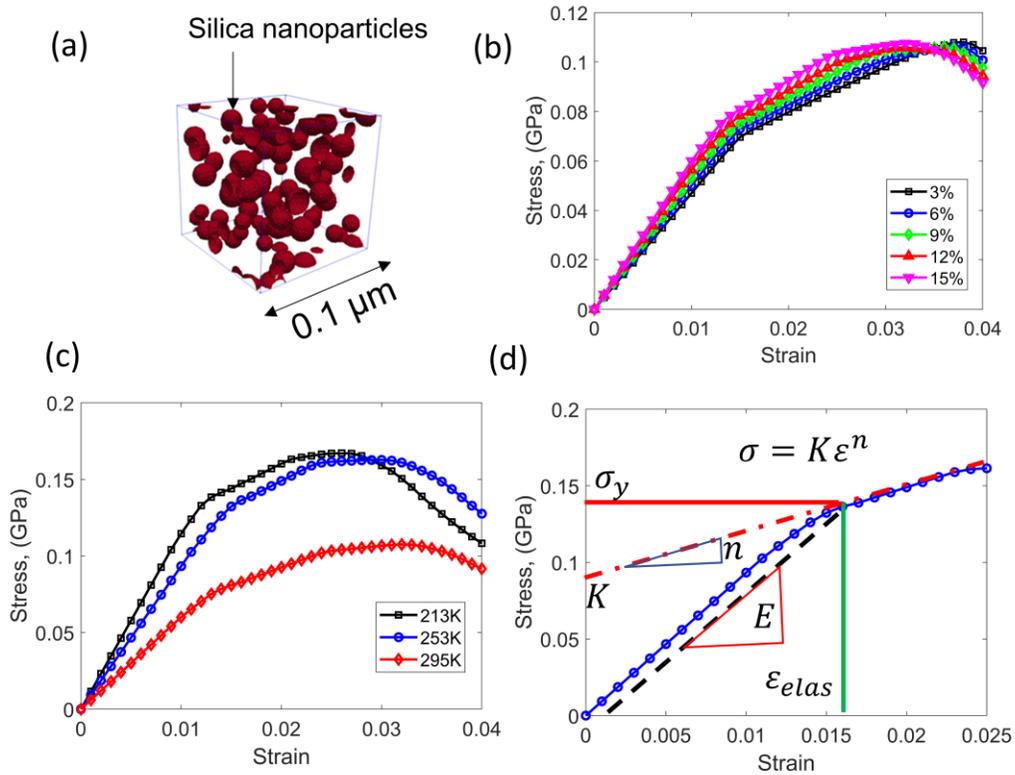

**Fig. 5**. (a) Silica reinforced epoxy structures with silica nanoparticles having average diameter of 20 nm, (b) stress-strain curve for various percentage of silica reinforcement in epoxy matrix, (c) temperature dependent stress-strain curve for 15% silica reinforced epoxy, (d) mechanistic feature extraction from a sample stress-strain curve.

**Microscale Unidirectional Composite Design**

For the microscale, we consider a unidirectional composite with carbon fiber as the reinforcing phase and the silica-reinforced epoxy as the matrix materials. The design parameters considered for the unidirectional microstructure are the volume fraction of the carbon fiber and the temperature. For the matrix phase, the design variables can be obtained from the nanoscale. Similar to the nanoscale, we start with the UD microstructure (see **Fig. 6(a)**) and use the SCA method to generate the stress-strain response. We apply the load in six different directions (3 normal and 3 shear direction) for the UD microstructures to obtain the stress-strain curve. Representative transverse and shear loading stress-strain curves are shown in **Fig. 6(b)** for fiber volume fractions of 50% and the silica fraction of 15% in the matrix materials. The generated stress-strain curves are further analyzed for the mechanistic feature extractions. As the UD microstructure comes from the yarn of the mesoscale woven structure, we are primarily interested in the elastic properties of this UD composite phase. For an orthotropic material like UD composites, only five elastic constants are sufficient instead of twenty-one. Longitudinal and transverse elastic and shear moduli and the poisons ratio are obtained by loading the UD composites in normal and shear directions. These elastic features are available in the composite knowledge database for further analysis.

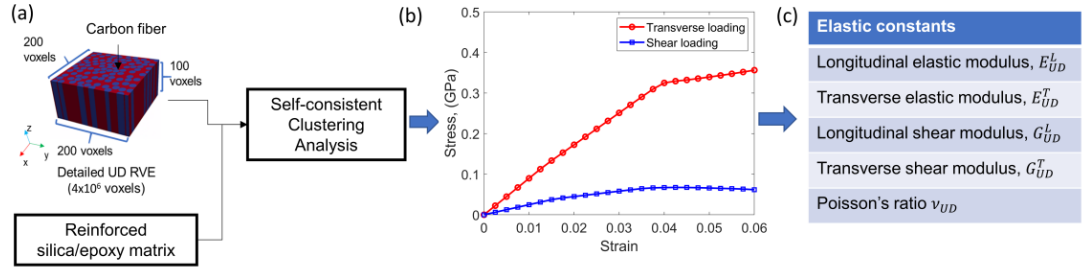

**Fig 6.** (a) Workflow for generating stress-strain curve using Self-consistent Clustering Analysis (SCA) method, (b) stress-strain response for transverse and shear loading of unidirectional composite with 50% fiber volume fraction and 15% silica reinforcement in the matrix, (c) extracted features from the stress-strain curve.

For the design analysis of a specific woven structure, the yarn properties such as volume fraction are important design parameters. This yarn is essentially a UD microstructure. Therefore, for immediate analysis of a mesoscale yarn, we need the UD data readily available. A mechanistic learning step utilizing a feed-forward neural network (FFNN) is used to map the microscale and nanoscale design variables (fiber and nanofiller volume fractions and temperature) to the elastic properties of the UD composite. The FFNN has a simple structure with two hidden layers (see **Fig. 7(a)**), and tangent sigmoid activation functions are used. Mean square error was set as the loss function. The hyperparameters were optimized using backward propagation. The number of neuron and layers are arbitrary here for demonstration purpose, and finding an optimum number of them are out of the scope of the current paper. With the ten neurons in each hidden layer, we can reach the coefficient of determination, $R^2$ value of 0.99, for all the training, validation, and testing set, which are set as 70%, 15%, and 15% of the total data, respectively.

After the neural network training, several cases are tested to see the accuracy of the trained model. **Fig. 7(b)** shows that the model is quite accurate in predicting all the five elastic constants when the input parameters are within the training range. This trained neural network is a part of the composite knowledge database that can be further used in the mesoscale woven composite design.

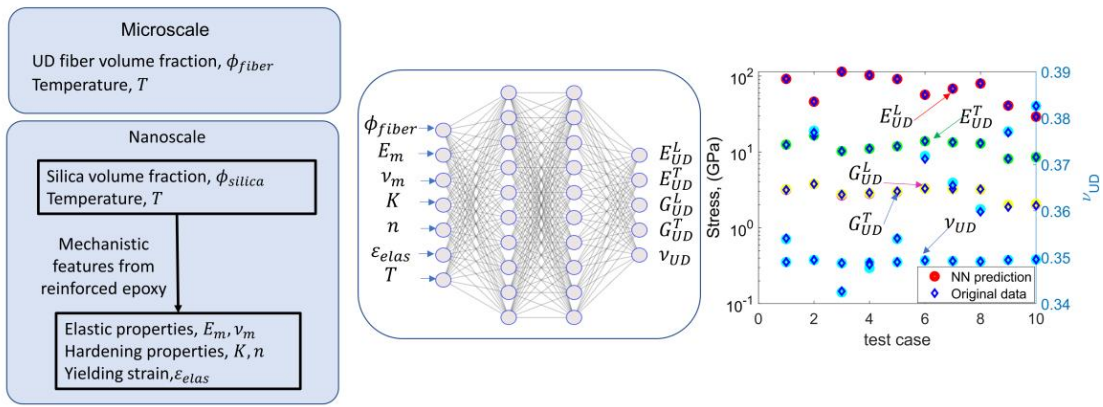

Fig. 7. (a) Mechanistic learning in microscale UD composite using a feed-forward neural network, (b) accuracy of neural network prediction for the elastic constants of the UD composites.

**Mesoscale Woven Composite Design**

For the mesoscale woven composite, the physical size of a composite is in the millimeter scale. A woven composite RVE and its typical physical size with different design parameters are shown in **Fig. 8(a)**. The woven composite is made of yarn, with carbon fibers, and the yarns are weaved into different patterns. In the polymer infiltration process, the polymer impregnates the fibers bundles and forms a microscale UD composite structure. For a woven-like structure, several design parameters can be established, such as the physical size of the RVE, the geometry of the weft and warp, yarn angle, yarn fiber volume fractions, etc. The matrix phase for the woven composite can still be the reinforced matrix that we already analyzed in the nanoscale design. The constituents of the woven structure are the UD structure as the yarn and the reinforced epoxy as the matrix. These material phases are already analyzed in the nano and microscale of the problem, and the relevant material properties are extracted and stored in the composite knowledge database. The woven design parameters can be studied in the woven scale, and the corresponding material properties (UD and reinforced matrix) are taken from the knowledge database. Once we define the design space for the woven composite and perform the analysis, mechanistic learning for the mesoscale can be set up where all the design parameters can be input together, and the response of the composite can be extracted. This composite response can be directly paired with a part-scale performance prediction and design step, as described in **Fig. 8(b)**.

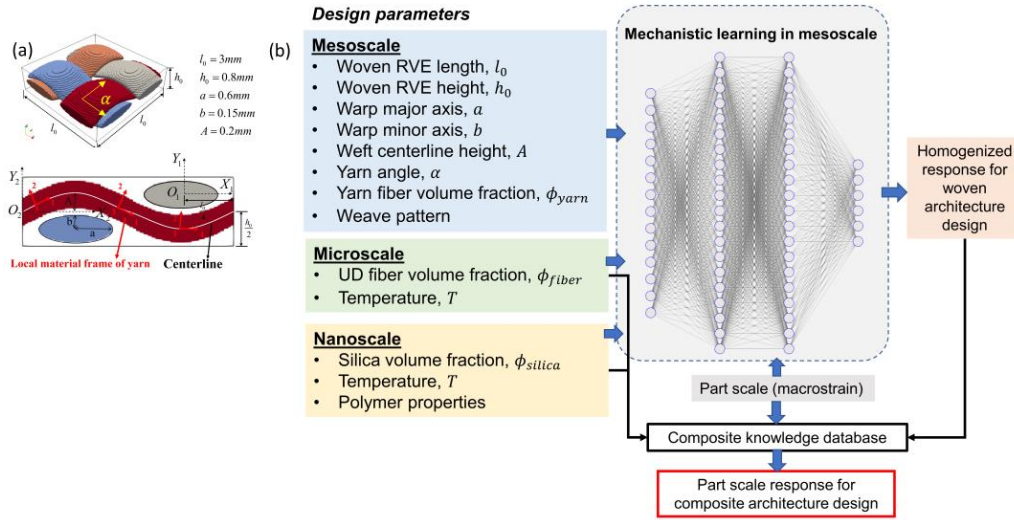

Fig 8. (a) Woven composite and its RVE with design parameters, (b) design parameters of composites at multiscale and composite knowledge database preparation for system and design.

Using the MDS approach, the multiple length scales of a composite can be analyzed systematically to evaluate the new composite deign and the performance prediction rapidly. In MDS, different length scales are analyzed that enrich the composite knowledge database, which makes it possible for rapid design iteration at the part scale level without re-evaluating the lower length scales again. Using these data science tools, such as neural networks, the hidden relationships among different design parameters and materials response are established that reveal the governing mechanism at different length scales for the composite materials.

**PERSPECTIVE AND FUTURE WORKS**

Polymer matrix-based composite materials have a promising future to replace conventional aerospace materials. Though polymer matrix composites have been studied in the last few decades, the mechanics and the modeling of this kind of heterogeneous materials remain challenging. Optimizing the design parameters for a composite structure at the multiple length scales can further ensure the safety operation that is essential for aerospace-grade materials. Also, superior properties and low materials cost can be achieved from an optimized design of the material systems at different length scales. Below we mention a few areas that require attention from the community to be addressed in the near future to make composite materials a sustainable solution for the high-tech such as aerospace industry.

- The design of a high-performance polymeric matrix relies on the analysis of its chemical compositions and structures. From the chemical structure point of view, critical building blocks like functional groups are well handled by existing data science tools. However, in terms of the microstructure of the polymeric matrix, it is still a challenging issue to consider polymer's entanglement, porosity, or crystallization, etc., in the property optimization. In addition, the interaction between the polymeric matrix and the reinforcing fillers brings another design parameter that affects the performance of the composite. Therefore, a comprehensive experimental dataset for the microstructure feature of the polymeric matrix is of great importance when building the MDS knowledge database.

- The multifunctionality of the composite materials can be incorporated by designing the materials from nano to the part scale. However, the experiments remain challenging to perform for the lower length scales, and a multi-scale analysis tool can provide insights on the design of composite materials for those scales. Though a few multi-scale tools are available, the field still lacks a real-time concurrent analysis tool that enables the designer to perform real-time analysis of a composite structure for design. By constructing the knowledge database at multiple scales, MDS can accelerate the design and analysis of a new structure significantly. The MDS framework can be further enhanced by developing new tools for its different modules. For instance, new reduced-order models can generate faster data, and new mechanisms can be identified using the mechanistic learning tool. This MDS-based multi-scale analysis tool can replace traditional multi-scale tools with a real-time prediction capability.

- In MDS, existing data science tools such as neural network, clustering, principal component analysis, etc., are used to analyze the data. These tools are mathematical tools that lack the physics of the problem. Integrating the right physical principles, these tools can be more beneficial for science and engineering problems. Further research is necessary to understand these data science tools and how they can be used for physical problems. One example can be the formulation of the hierarchical deep learning neural network FEM (HiDeNN-FEM) method [65], where the neural network is used to construct the shape function of the finite element. Using the universal approximation capabilities of the neural network, h-, p-, hp- adaptivity of the finite element can be achieved.

- The key benefit of the MDS is that once the knowledge database is created, it can be used for the further design and analysis of a new system. This capability enables the materials designer to analyze a new materials system and design rapidly.

A design iteration loop can be used for the materials property optimization and the structure level topological optimization to harness the maximum potential of the newly designed material systems. A design tool can be plugged in with the MDS knowledge database that allows the designer to perform real-time design and analysis.

## ACKNOWLEDGMENT

Satyajit Mojumder and Wing Kam Liu gratefully acknowledge the financial support provided by the Air Force Office of Scientific Research (FA9550-18-1-0381). Lei Tao and Ying Li gratefully acknowledge financial support from the Air Force Office of Scientific Research through the Air Force's Young Investigator Research Program (FA9550-20-1-0183; Program Manager: Dr. Ming-Jen Pan).

## FUNDING

Funding for this work is provided by the Air Force Office of Scientific Research through grants no. FA9550-18-1-0381 and FA9550-20-1-0183.

**Figure Captions List**

Fig. 1     Aerospace composite materials and its structure in nanoscale as reinforced matrix, microscale as unidirectional structure, mesoscale as woven structure and a part scale three point bending sample

Fig. 2     Mechanistic data science framework and its six components to create knowledge database for composite design, analysis and performance optimization

Fig. 3     Discovery of high temperature polymers. A DNN machine learning model is first trained on a dataset of ~7000 existing polymers. The using the model to screen 1 million hypothetical polymers to discover ~65,000 promising high temperature structures whose $T_g$ > 200°C.

Fig. 4     The forward prediction of polymer properties and the inverse design of polymer structures. Variational autoencoder (VAE) and generative adversarial networks (GAN) generates new structures whose properties can be evaluated by the forward prediction machine learning model

Fig. 5     (a) Silica reinforced epoxy structures with silica nanoparticles having average diameter of 20 nm, (b) stress-strain curve for various percentage of silica reinforcement in epoxy matrix, (c) temperature dependent stress-strain curve for 15% silica reinforced epoxy, (d)mechanistic feature extraction from a sample stress-strain curve

Fig. 6       (a) Workflow for generating stress-strain curve using Self-consistent Clustering Analysis (SCA) method, (b) stress-strain response for transverse and shear loading of unidirectional composite with 50% fiber volume fraction and 15% silica reinforcement in the matrix, (c) extracted features from the stress-strain curve

Fig. 7       (a) Mechanistic learning in microscale UD composite using a feed-forward neural network, (b) accuracy of neural network prediction for the elastic constants of the UD composites

Fig. 8       (a) Woven composite and RVE and design parameters, (b) design parameters of composites at multi-scale and composite knowledge database preparation for system and design

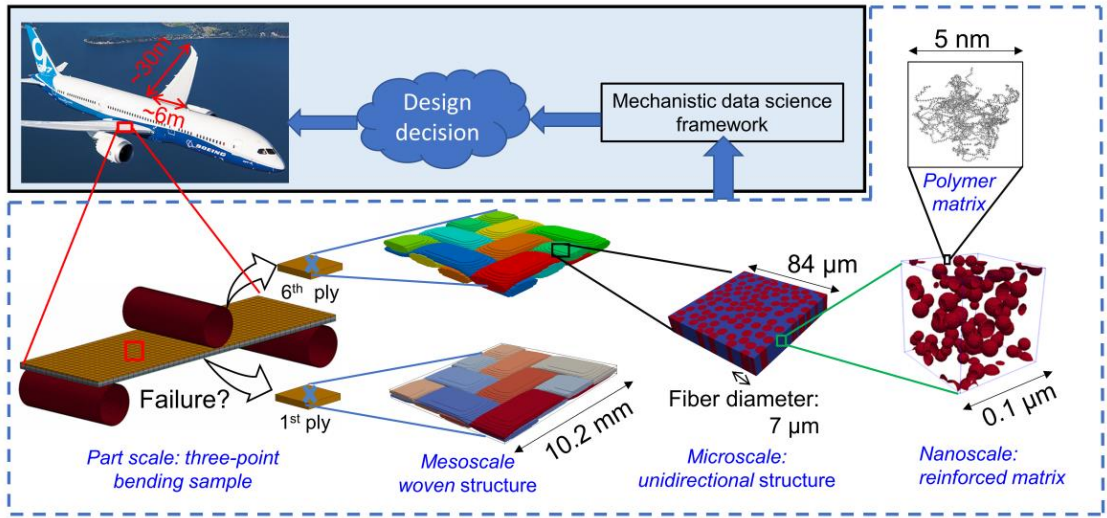

Fig. 1. Aerospace composite materials and its structure in nanoscale as reinforced matrix, microscale as unidirectional structure, mesoscale as woven structure and a part scale three point bending sample

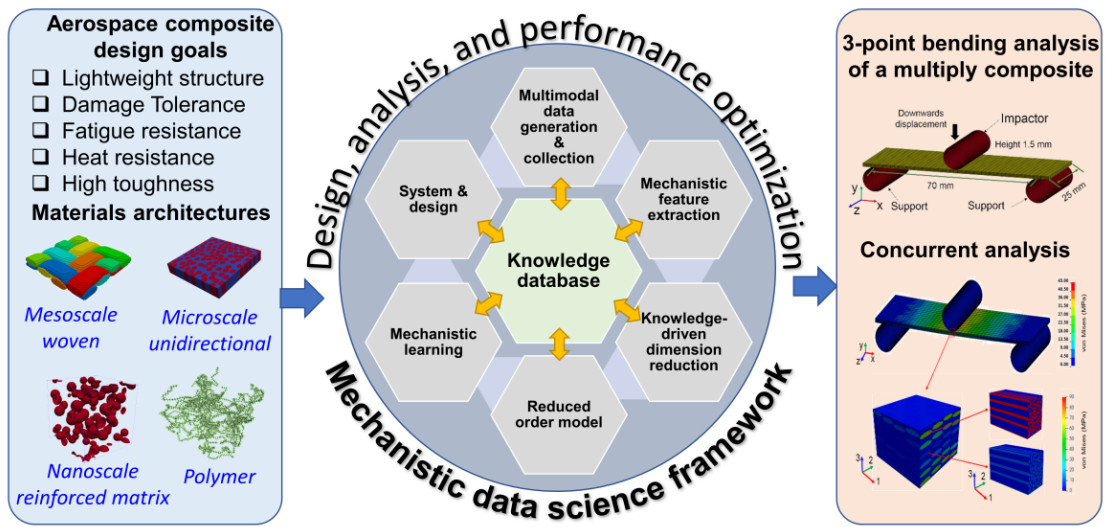

Fig 2. Mechanistic data science framework and its six components to create knowledge database for composite design, analysis and performance optimization

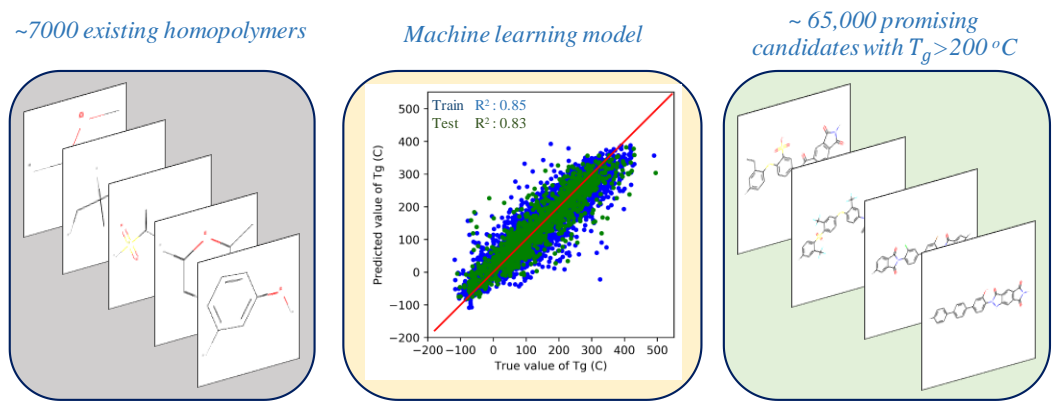

Fig 3. Discovery of high temperature polymers. A DNN machine learning model is first trained on a dataset of ~7000 existing polymers. The using the model to screen 1 million hypothetical polymers to discover ~65,000 promising high temperature structures whose $T_g$ > 200°C.

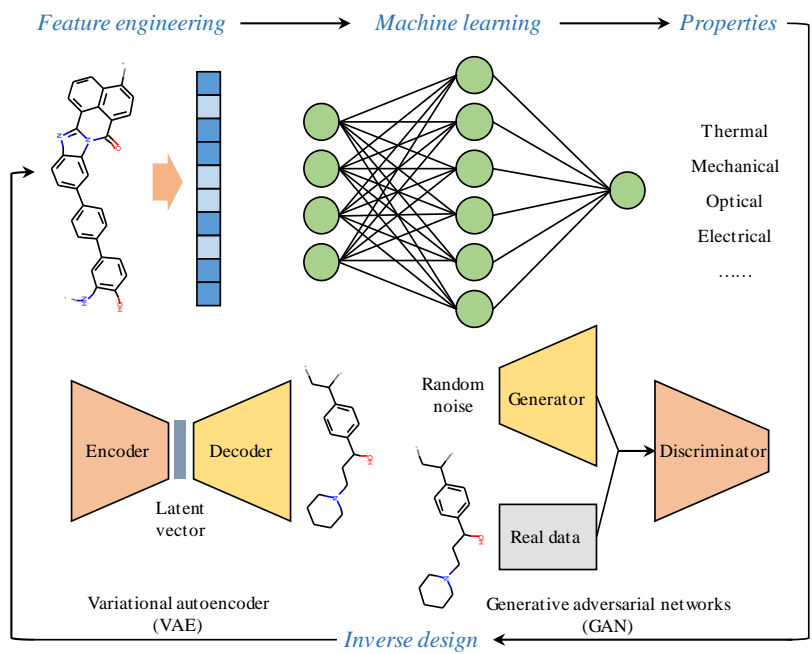

Fig 4. The forward prediction of polymer properties and the inverse design of polymer structures. Variational autoencoder (VAE) and generative adversarial networks (GAN) generates new structures whose properties can be evaluated by the forward prediction machine learning model.

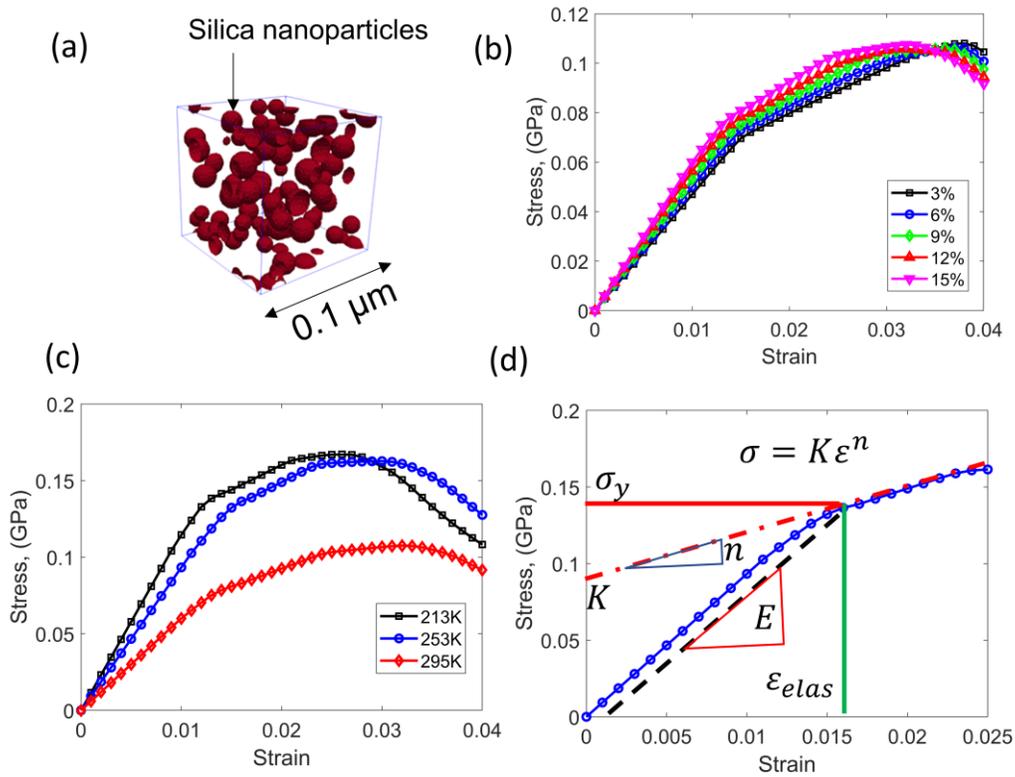

Fig. 5. (a) Silica reinforced epoxy structures with silica nanoparticles having average diameter of 20 nm, (b) stress-strain curve for various percentage of silica reinforcement in epoxy matrix, (c) temperature dependent stress-strain curve for 15% silica reinforced epoxy, (d)mechanistic feature extraction from a sample stress-strain curve.

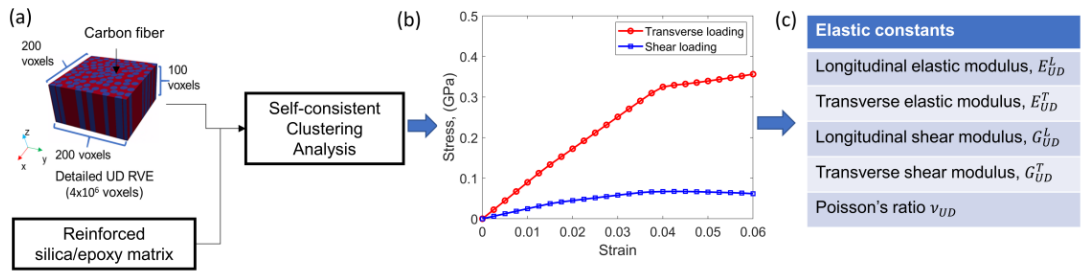

Fig 6. (a) Workflow for generating stress-strain curve using Self-consistent Clustering Analysis (SCA) method, (b) stress-strain response for transverse and shear loading of unidirectional composite with 50% fiber volume fraction and 15% silica reinforcement in the matrix, (c) extracted features from the stress-strain curve.

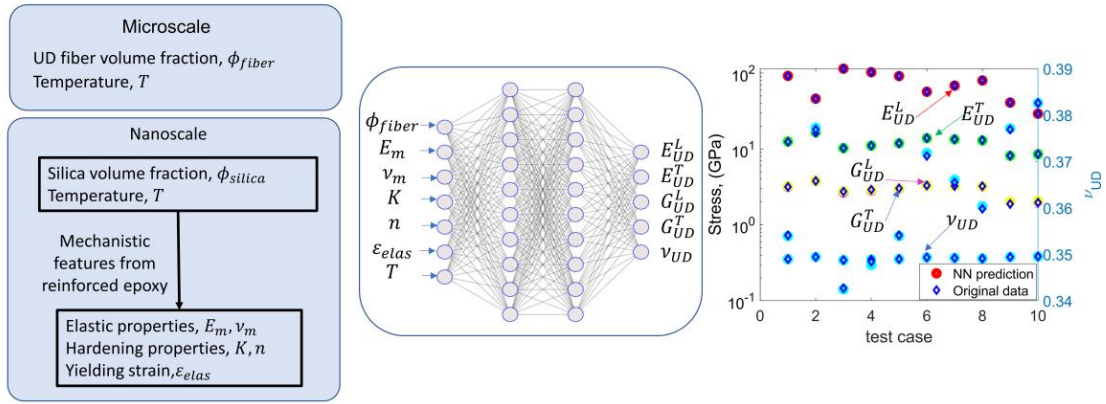

Fig. 7. (a) Mechanistic learning in microscale UD composite using a feed-forward neural network, (b) accuracy of neural network prediction for the elastic constants of the UD composites.

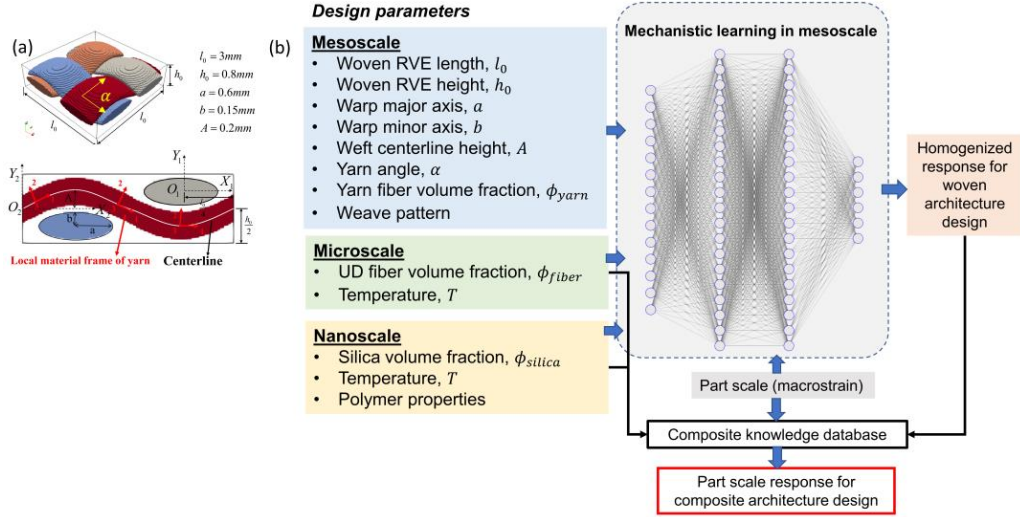

Fig 8. (a) Woven composite and RVE and design parameters, (b) design parameters of composites at multiscale and composite knowledge database preparation for system and design.